\documentstyle[aps,multicol,amssymb,graphicx]{revtex}

\begin{document}
\draft
\preprint{}
\title{\bf Comment on ``Critique of $q$-entropy for thermal statistics" by M. Nauenberg}
\author{C. Tsallis \thanks{tsallis@cbpf.br}
 }
\address {Centro Brasileiro de Pesquisas Fisicas, Rua Xavier Sigaud 150, 22290-180
Rio de Janeiro - RJ, Brazil}
\date{\today}
\maketitle
%\begin{abstract}

%\end{abstract}

{\bf Pacs Numbers:} 05.70.-a, 05.20.-y, 05.90.+m

\vspace{1.0cm}

%\section{Introduction}
%\label{sec:intro}

\begin{multicols}{2}

It was recently published  \cite{nauenberg} a quite long list of objections about the physical validity for thermal statistics of the theory sometimes referred to in the literature as {\it nonextensive statistical mechanics}. This generalization of Boltzmann-Gibbs (BG) statistical mechanics is based on the following expression for the entropy:   
\begin{equation}
S_q= k\frac{1- \sum_{i=1}^Wp_i^q}{q-1} \;\;\;(q \in {\cal R}; S_1=S_{BG} \equiv -k\sum_{i=1}^W p_i \ln p_i) \;. 
\end{equation}
The author of \cite{nauenberg} already presented orally the essence of his arguments in 1993  during a scientific meeting  \cite{buenosaires}. I am replying now simultaneously to the just cited paper, as well as to the 1993 objections (essentially, the violation of ``fundamental thermodynamic concepts", as stated in the Abstract of \cite{nauenberg}). The list of objections and critical statements being extremely long, it is perhaps not really necessary at the present stage to reply to all the points. For time and space economy, I will therefore address here only a few selected points, hopefully the most relevant ones, physically and/or logically speaking.   \\

\noindent
{\it About the nonextensivity of the entropy $S_q$:}\\

The entropy $S_q$ is nonextensive for independent systems (see Eq. (6) of \cite{nauenberg}), {\it which by no means implies that it cannot be extensive in the presence of correlations at all scales}. Nowhere in \cite{nauenberg} is there clear evidence of taking this fact into account in what concerns the validity of the $q$-thermostatistics. It is nevertheless of crucial importance, as we illustrate now for the simple case of equiprobability (i.e., $p_i=1/W, \;\forall i$). In such simple situation, Eq. (1) becomes
\begin{equation}
S_q=k \ln_qW \;\;\;(\ln_q x\equiv (x^{1-q}-1)/(1-q); \ln_1 x=\ln x)\;.
\end{equation} 
If a system constituted by $N$ elements is such that it can be divided into two or more essentially independent subsystems (e.g., independent coins or dices, or spins interacting through short-range coupling), we generically have $W \sim \mu^N$ ($\mu>1$). Consequently, $S_q/k \sim \ln_q \mu^N$. There is an unique value of $q$, namely $q=1$, for which we obtain the usual result $S_q \propto N$. But if the system is such that we have $W \sim N^\rho$ ($\rho>0$), then $S_q/k \sim \ln_q N^\rho$. Once again, there is a unique value of $q$, namely $q=1-1/\rho$ for which, $S_q \propto N$. This fact is well known to many scientists working on nonextensive statistical mechanics, and has been published in the special volume dedicated to the subject indicated in Refs. [5, 6, 14] of paper \cite{nauenberg}.  The same property holds in fact for $S_{\sigma(q)}$, $\sigma(q)$ being any smooth function of $q$ such that $\sigma(1)=1$ (e.g., $\sigma=1/q$, or $\sigma=2-q$). For the correlated case, we have $S_{\sigma(q)} \propto N$ only for $q$ satisfying $(1-\sigma(q)) \rho=1$. The relevance of this property ($S \propto N$) for thermodynamics needs, we believe, no further comments.\\

\noindent
{\it About the concept of ``weak coupling" in} \cite{nauenberg}{\it :}\\

Much of the criticism in \cite{nauenberg} involves the concept of ``weak coupling". To make this point clear through an illustration, let us think of the ground state of a Hamiltonian many-body classical system whose elements are localized on a $d$-dimensional lattice and have two-body interactions among them. Let us further assume that the (attractive) coupling constant  is given by $C_{ij}=-c/r_{ij}^\alpha$ ($c>0$, $\alpha \ge 0$, and $r_{ij}=1, ...$). The potential energy $U(N)$ per particle generically satisfies $U(N)/N \propto -c\sum_{i \ne j}1/r_{ij}^\alpha \simeq -c \int_1^{N^{1/d}} dr\, r^{d-1}\, r^{-\alpha} \propto -c \frac{N^{1-\alpha/d}-1}{1-\alpha/d}  $ . Therefore, for $\alpha/d >1$ ({\it short-range} interactions in the present context), we have that $lim_{N \to\infty} U(N)/N$ is finite, and BG statistical mechanics certainly provides the appropriate answer for the stationary state (thermal equilibrium) of the system. In this case, all the usual prescriptions of thermodynamics are satisfied, as well known \cite{fisheretal}. If the interactions are, however, {\it long-range} (i.e., $0 \le \alpha/d \le 1$), then $lim_{N \to\infty} U(N)/N$ diverges, and the case needs further discussion. It might well happen that, dynamically speaking, the $N\to\infty$ and the $t\to\infty$ limits do not commute. If so, only the $\lim_{N\to\infty}\lim_{t\to\infty}$ ordering corresponds to the BG stationary state, whereas the opposite ordering,  $\lim_{t\to\infty}\lim_{N\to\infty}$, might be a complex one, {\it different} from the BG state, and in some occasions possibly related to the one obtained within the $q$-formalism. It is clear then that, if we have long-range interactions and $N>>1$ (say of the order of the Avogadro number), it might very well happen that the BG equilibrium is physically inaccessible, and the only physically relevant stationary or quasi-stationary (metastable) state is a non-Gibbsian one. Such situation is indeed found in \cite{rapisarda}, as discussed below.   

We can now address the manner used in \cite{nauenberg} to refer to ``weak coupling". It applies essentially in the simple manner stated in \cite{nauenberg} only for $\alpha/d>1$, being {\it conceptually much more subtle} for $0 \le \alpha/d \le1$. For example, if  $0 \le \alpha/d <1$, $U(N)/N$ {\it diverges} as $N^{1-\alpha/d}$ ($N \to\infty$) for {\it any} nonvanishing value of $c$, even for $c$ corresponding to ... $10^{-10}eV$!   Consistently, the {\it generic} use, without further considerations (such as  the $(N,t) \to (\infty,\infty)$ limits, and the range of $\alpha/d$), of relations such as Eqs. (5) and (7) of \cite{nauenberg} seems irreducibly unjustified; as they stand, they trivially yield to no other possibility than $q=1$. In fact, this point has already been transparently addressed by Fermi in  1936 \cite{fermi}.  \\

\noindent
{\it About the determination of the value of $q$ for a given system:}\\

The entropic parameter $q$ is referred in \cite{nauenberg} as an ``undetermined parameter". Moreover, the author claims having proved that ``$q$ must be a {\it universal} constant, just like the Boltzmann constant $k$...". I have difficulty in unambiguously finding in the paper whether this kind of statement would only apply to Hamiltonian systems, or perhaps also to dissipative ones; to systems whose phase space is high-dimensional, or perhaps also to the low-dimensional ones. By ``undetermined", it remains not totally clear whether the expression is used in the sense that $q$ is ``undeterminable", or in the sense of ``not yet determined". However, if we put all this together, one might suspect that what is claimed in \cite{nauenberg} is that it {\it can} be determined from first principles, and that the author has determined it to necessarily be $q=1$. 

To make this point transparent, we may illustrate the factual {\it nonuniversality} of $q$ by addressing the logistic-like family of maps $x_{t+1}=1-a|x_t|^z$, whose usefulness in physics can hardly be contested (at least for $z=2$). As conjectured since 1997 \cite{TMP}, numerically exhibited in many occasions (e.g., in \cite{uriel,lyra,lyramoura,ernestogarin}),  and analytically proved recently \cite{fulvioalberto,fulvioalberto2} on renormalization group grounds, {\it $q$ does depend on $z$}, and is therefore {\it not} universal, in neat contrast with what is claimed in \cite{nauenberg}. Its value for $z=2$ (i.e., the standard logistic map), as given by the sensitivity to the initial conditions, is $q=0.244487...$ at the edge of chaos (e.g., $a=1.401155...$), whereas it is $q=1$ for all values of $a$ for which the Lyapunov exponent is positive (e.g., for $a=2$). We have illustrated the nonuniversality of $q$ for nonlinear dynamical systems with its value at the edge of chaos of the logistic map. It is perhaps worthy to notice that, since it has been proved to be analytically related to the Feigenbaum universal constant $\alpha_F$ [$1/(1-q)=\ln \alpha_F/\ln 2$], and since this constant is already known with not less than 1018 digits, we actually know this particular value of $q$ with the same number of digits. Such a precision is self-explanatory with regard to the fact that $q$ {\it can} be determined from first principles and that it {\it can} be different from unity (see also \cite{scienceplastino,sciencelatora}). 

A second illustration of the {\it nonuniversality} of $q$ can be found in the three-component Lotka-Volterra model  in a $d$-dimensional hypercubic lattice \cite{LV}. This illustration is quite interesting because this model is a {\it many-body} problem. The corresponding growth of droplets has been shown to yield, through imposition of the finiteness of the entropy production per unit time, $q=1-1/d$ for $d=1,2$ \cite{astero}. This law has also been checked for $d=3,4$ \cite{celia}. It is of a rather simple nature, essentially  related to the fact that the growth of the bulk regions of this specific model is characterized by the droplet linear size linearly increasing with time \cite{celia}. A similar law is obtained for a quite different model, namely a Boltzmann $d$-dimensional lattice model for the incompressible fluid Navier-Stokes equations. Indeed, an unique entropy, namely $S_q$, with an unique value of $q$, is mandated by the imposition of the most basic Galilean invariance of the equations. For the single-speed single-mass model it is 
$q=1-2/d$ \cite{bogho1}. For more sophisticated models $q$ is determined by a transcendental equation \cite{bogho2}. These examples show how the entropy $S_q$ enables to put {\it on equal grounds} situations that are physically quite disparate. 

Given the preceding illustrations of dissipative systems, and many others existing in the literature, it could hardly be a big surprise if, {\it also} for many-body Hamiltonian systems, $q$ turned out to be a nonuniversal index essentially characterizing what we may consider as nonextensivity universality classes (in total analogy with the universality classes that emerge in the theory of critical phenomena). More precisely, one expects $q=1$ for short-range interactions ($\alpha/d >1$ in the example we used earlier), and $q$ depending on $(d,\alpha)$ (perhaps only on $\alpha/d$) for long-range interactions (i.e., $0 \le \alpha/d <1$), in the physically most important ordering $\lim_{t\to\infty} \lim_{N\to\infty}$. Although expected, the uncontestable evidence has not yet been provided. It is not hard for the reader to imagine the analytic and computational difficulties that are involved. However, suggestive results are accumulating which point towards the applicability of nonextensive statistical mechanics for such long-range Hamiltonian systems. Although we shall later come back onto this problem, let us already mention the following points. 

(i) The one-body marginal distribution of velocities during the well known longstanding quasi-stationary (metastable) state of  the isolated classical inertial $XY$ ferromagnetically coupled rotators localized on a $d$-dimensional lattice can be anomalous (i.e., non-Maxwellian). Indeed, it approaches, for a non-zero-measure class of initial conditions of the
$\alpha=0 \;(\forall d)$ model and not too high velocities, a $q$-exponential distribution (we remind that $e_q^x \equiv [1+(1-q)x]^{1/(1-q)}$, hence $e_1^x=e^x$) with $q>1$ \cite{rapisarda}. {\it If} the energy distribution followed BG statistics, the one-body marginal distribution of velocities {\it ought} to be quasi-Maxwellian ({\it strictly} Maxwellian in the $N \to \infty$ limit since then the microcanonical-ensemble necessary cutoff in velocities diverges), but  it is {\it not} .  As specifically discussed in \cite{rapisarda}, the numerical results are incompatible with BG statistics. However, they do not yet prove that the one-body distribution of velocities  precisely is, for the canonical ensemble, the one predicted by nonextensive statistics. Indeed, considering the appropriate limit $(N,M,N/M) \to (\infty,\infty,\infty)$ ($N$ being the number of rotators of the isolated system, and $M$ being that of a relatively small subsystem of it) is crucial. Work along this line is in progress.   

(ii) In the same model, at high total energy, the largest Lyapunov exponent vanishes like $1/N^\kappa$ where $\kappa$ depends on $\alpha/d$ \cite{celiaconstantino,giansanti}. Also during the lonstanding state, the largest Lyapunov exponent vanishes, this time like $1/N^{\kappa/3}$ \cite{cabral}. It is clear that, with a vanishing Lyapunov spectrum, the system will be seriously prevented from satisfying Boltzmann's ``molecular chaos hypothesis", hence the ``equal probability" occupation of phase space. 

(iii) In the longstanding regime of the $\alpha=0\;(\forall d)$ model, there is {\it aging} \cite{montemurroetal}, something which is {\it totally incompatible with the usual notion of thermal equilibrium}. The correlation functions depend on the ``waiting time", and are in all cases given by $q$-exponential functions. Even at high total energy, where the one-body distribution of velocities is Maxwellian, and where there is {\it no} aging, the time correlation functions are {\it still}  given by $q$-exponentials with $q>1$, instead of exponentials, which is the standard expectation in BG statistics.    

(iv) The temperature ($\propto$ mean kinetic energy per particle) relaxes, after the quasistationary state observed in the one-dimensional $0\le \alpha<1$ model, towards the BG temperature through a $q$-exponential function with $q>1$ \cite{rapisardarecente}.  

(v) In Lennard-Jones clusters of up to $N=14$ atoms, the distribution of the number of links per site has been numerically computed \cite{doye}, where two local minima of the many-body potential energy are ``linked" if and only if they are separated by no more than one saddle-point. This distribution is a $q$-exponential with $q \simeq 2$, as can be checked through direct fitting. The possible connection with our present discussion comes from the fact that the average diameter of the cluster is (in units of atomic size) of the order of $14^{1/3} \simeq  
2.4$. Consequently, although the Lennard-Jones interaction is {\it not} a long-range one thermodynamically speaking (indeed, $\alpha/d=6/3=2>1$), it can effectively be considered as such for small clusters, since all the atoms substantially interact with all the others.  

(vi) The distribution of the number of links per node for the Albert-Barabasi growth model \cite{barabasi} yielding scale-free networks is analytically established to be, in the stationary state, a $q$-exponential with $q=[2m(2-r)+1-p-r]/[m(3-2r)+1-p-r] \ge 1$, where $(m,p,r)$ are microscopic parameters of the model. If we associate to this network an $\alpha=0$ interaction per link, the just mentioned  distribution also represents the distribution of energies per node. Although this is not the same distribution as that of the energy of microscopic states associated with a Hamiltonian, it is neither very far from it.

(vii) Although not being  many-body problems, let us mention at this point some results that have been obtained with the $d=2$ standard map and with a $d=4$ set of two coupled standard maps. Both systems are {\it conservative} and {\it simplectic}, having therefore the dynamical setup of a standard Hamiltonian. The $d=4$  system has  Arnold diffusion as soon as the nonlinear coupling constant $a$ is different from zero;  this guarantees a chaotic sea which is singly connected in phase space (we may say that $a_c=0$). The structure is more complex for the $d=2$ case because no such diffusion is present; consistently, unless $a$ is sufficiently large, disconnected chaotic ``lakes" are present in the phase space; below $a_c = 0.97...$, closed KAM regions emerge in the problem. The remark that we wish to do here is that, in strong analogy with the many-body long-range Hamiltonian cases we have been discussing, both the $d=2$ and the $d=4$ maps present a longstanding quasistationary states {\it before} crossing over to the stationary ones. The crossover time $t_{crossover}$ diverges when $a$ approaches $a_c$ from above. This is very similar to what happens with the above $(d,\alpha)$  Hamiltonian, for which strong numerical evidence exists \cite{rapisarda,cabral,giansantietal} suggesting that $t_{crossover}$ diverges as $(N^{1-\alpha/d}-1)/(1-\alpha/d)$ when $N \to\infty$.     

Although none of the (seven) factual arguments that we have just presented constitutes a proof, the set of them does provide,  in our undestanding, a quite strong suggestion that the longstanding quasistationary states existing in long-range many-body Hamiltonians might be intimately connected to the nonextensive statistics, with $q$ depending on basic model parameters such as $d$ and $\alpha$. The entropic index $q$ would then characterize {\it universality classes of nonextensivity}, the most famous of them being naturally the $q=1$, extensive, universality class. Such viewpoint is also consistent with the discussion about non-Gibbsian statistics presented in \cite{sokal}. Last but by no means least, it is consistent with Einstein's 1910 criticism \cite{einstein} of the Boltzmann principle $S=k \ln W$ (lengthily commented in Ref. [6] of \cite{nauenberg}).\\   

\noindent
{\it About thermal contact between systems with different values of $q$ and the $0^{th}$ principle of thermodynamics:}\\

We focus now on a strong and crucial statement in \cite{nauenberg}, namely "... a Boltzmann-Gibbs thermometer would not be able to measure the temperature of a $q$-entropic system, and the laws of thermodynamics would therefore fail to have general validity." \cite{nauenberg}. We shall present here the results \cite{moyanobaldovin} of molecular-dynamical simulations (using {\it only} $F=ma$ as microscopic dynamics) which will {\it precisely exhibit what is claimed in} \cite{nauenberg} {\it to be impossible}. We shall illustrate this with the isolated $\alpha=0$ model of planar rotators, and proceed through two steps. 

We first show (Fig. 1) how the ``temperature" (defined as twice the instantaneous kinetic energy per particle) of a relatively small part of a large system relaxes onto the ``temperature" of the large system {\it while this is in the quasistationary regime} (where the system has been definitely shown to be {\it non}-Boltzmannian, and where it might well be described by the $q$-statistics). We verify that the rest of the system acts for a generic small part of itself as a ``thermostat", {\it in total analogy with what happens in BG thermal equilibrium}. This is quite remarkable if we think that the system is in a state {\it so different from thermal equilibrium that it even has aging!} 

We then show (Fig. 2) how a BG thermometer (its internal degrees of freedom are those of  {\it first-neighbor-coupled} inertial rotators, hence definitively a $q=1$ system) {\it does} measure the ``temperature" of the {\it infinitely-range-coupled} inertial rotators during their quasistationary state, hence where the statistics is definitely {\it non}-Boltzmannian. At the light of this evidence, it appears that the $0^{th}$ principle of thermodynamics is even more general than the already important role that BG statistical mechanics reserves for it. Naturally, the fluctuations that we observe in both figures are expected to disappear in the $(N,M,N/M) \to (\infty,\infty,\infty)$ limit. 

The facts that we have mentioned up to this point heavily disqualify the essence of the critique presented in \cite{nauenberg}.  I believe, nevertheless, that it is instructive to further analyze it. \\  

\noindent
{\it About the existing mathematical foundations of nonextensive statistical mechanics:}\\

It is essentially claimed in \cite{nauenberg} that it can be proved, from the very foundations of statistical mechanics, that the only physically admissible one is that of BG. It is however intriguing how such a strong statement may be done without clearly pointing the mathematical errors that should then exist in the available proofs of the $q$-exponential distribution. Such proofs have been provided by Abe and Rajagopal \cite{proof1,proof2,proof3,scienceabe}; they are multiple, mutually consistent, and generalize the well known proofs done, for BG statistics, by Darwin-Fowler (in 1922), Khinchin (in 1949) and Balian-Balazs (in 1987), respectively using the steepest-descent method \cite{proof1}, the laws of large numbers \cite{proof2}, and the counting for the microcanonical ensemble \cite{proof3}. All these proofs are ignored in \cite{nauenberg}. 
The critique therein developed outcomes severely diminished. 

Similarly, no mention at all is made in \cite{nauenberg} of the $q$-generalizations of Shannon 1948 theorem, and of Khinchin 1953 theorem, which are universally considered as part of the foundations of BG statistical mechanics since they prove under what conditions $S_{BG}$ is unique. These two $q$-generalizations \cite{theorem1,theorem2} analogously  exhibit the necessary and sufficient conditions associated with the uniqueness of $S_q$. 

Finally, no mention at all is made of the fact that $S_q$ ($\forall q >0$) shares with $S_{BG}$ three remarkable mathematical properties that are quite hard to satisfy, {\it especially  simultaneously}. These three properties are {\it concavity} (Ref. [1] of \cite{nauenberg}), {\it stability} \cite{stability}, and {\it finiteness of entropy production per unit time} (see \cite{production}, among others). The difficulty of having such agreable mathematical features can be measured by the fact that Renyi entropy (Eq. (19) of \cite{nauenberg}), for instance, satisfies {\it none} of them for abitrary $q>0$. 

It is perhaps for not paying due attention to all these theorems that the cyclic argument involving Eqs. (22-26) of \cite{nauenberg} has been included in the critique. Indeed, that argument uses Eq. (22) to ``prove" Eq. (26). Such a consistency can hardly be considered as surprising since the distribution in Eq. (22) is currently established {\it precisely using the BG entropy}, i.e., the form of Eq. (26). By the way, immediately after Eq. (26) we read ``provided $f(1)=f(0)=0$, which corresponds to the requirement that the entropy vanishes at $T=0$". It is in fact only $f(1)=0$ which is related to the vanishing entropy at $T=0$. The property $f(0)=0$ has in general nothing to do with it; it is instead related to the {\it expansibility} of the entropy, i.e., the fact that $S(p_1,p_2,...,p_W,0)=S(p_1,p_2,...,p_W)$.\\

\noindent
{\it About existing exact solutions of anomalous Fokker-Planck and Langevin equations:}\\

The standard $d=1$ Langevin equation (with a drift coefficient $\gamma$ and an additive noise), and the standard $d=1$ Fokker-Planck equation admit as exact solutions the Gaussian distribution, and are usually considered as paradigmatic {\it mesoscopic} descriptions associated with BG statistical mechanics. They can be naturally generalized by also including a multiplicative noise (with amplitude $M$) in the Langevin equation, and by considering the so called ``porous medium equation", i.e., a nonlinear Fokker-Planck equation where the Laplacian operator applies to the power $\nu$ of the distribution. The {\it exact} solutions of these two nontrivial (nonlinear) equations are $q$-Gaussians, with $q=(3M+\gamma)/(M+\gamma) \ge 1$ for the former \cite{multiplicative}, and $q=2-\nu <3$ for the latter \cite{nonlinear}. These suggestive mathematical facts are ignored in \cite{nauenberg}.  \\

\noindent  
{\it Miscellanous}\\

The precise formulation of nonextensive statistical mechanics has, since 1988, evolved along time in what concerns the way of imposing the auxiliary constraints under which $S_q$ is optimized (see Refs. [1-3] of \cite{nauenberg}). The paradigmatic case occurs for the canonical ensemble, where one must decide how to generalize the traditional energy constraint. The correct manner is nowadays accepted to be that indicated in Ref. [3] of \cite{nauenberg}, i.e., Eq. (3) of \cite{nauenberg}, namely
\begin{equation}
\frac{\sum_{i=1}^W p_i^q \epsilon_i}{\sum_{j=1}^W p_j^q }=U_q
\end{equation}
This particular writing of the energy constraint has various interesting features. Let us mention here three of them (further convenient features can be found in Ref. [3] of \cite{nauenberg}).

(i) It is precisely this form which emerges naturally within the steepest-descent proof \cite{proof1} of the $q$-statistics. It is a trivial consequence of the fact that $d e_q^x /dx=(e_q^x)^q$.   

(ii) This particular form makes  the theory to be, in what concerns the energy distribution, valid up to a {\it single} value of $q$, namely {\it precisely} that determined by the trivial constraint $\sum_{i=1}^W p_i=1$. Let us illustrate this in the continuum limit, for a typical example where the density of states $g(\epsilon) \propto \epsilon^\gamma$ for $\epsilon \to\infty$ ($\gamma \in {\cal R}$). Since we wish $p(\epsilon)$ to be normalizable, we must impose that $\int_{constant}^\infty d\epsilon \, g(\epsilon)\, p(\epsilon)$. Since $p(\epsilon) \propto \epsilon^{1/(1-q)}$   for $\epsilon \to\infty$, it must be $\gamma + 1/(1-q) >-1$, hence 
\begin{equation}
q<(2+\gamma)/(1+\gamma) \;,
\end{equation} 
($q<2$ for the simple case of an asymptotically constant density of states, i.e., $\gamma=0$). The finiteness of constraint (3) imposes $\int_{constant}^\infty d\epsilon \, g(\epsilon)\, \epsilon\, [p(\epsilon)]^q$ to be finite, which, interestingly enough, yields the {\it same} upper bound as before, namely Eq. (4). In other words, this makes the theory to have {\it both} constraints (norm and energy) mathematically well defined (i.e., given by {\it finite} numbers) {\it all the way up to a single upper bound for $q$}.

(iii) This structure (based on escort distributions \cite{escort}) for the energy constraint allows the construction of a quite general entropic form \cite{tsallissouza} which is extremized by  the Beck-Cohen superstatistics \cite{beckcohen}, and which, quite remarkably, is stable \cite{souzatsallis} (like $S_q$, and in variance with Renyi entropy).\\

Since 1988, many applications have been proposed for the nonextensive statistics. Some of these have been elaborated within the 1988 way of writing the energy constraint (Ref. [1] of \cite{nauenberg}), others have been elaborated within the 1991 way of writing this constraint (Ref. [2] of \cite{nauenberg}), and finally others with the 1998 way (Ref. [3] of \cite{nauenberg}). It is unfortunate that the 1998 way was not found from the very beginning in 1988, but this is the way it did happen (for a variety of reasons that are essentially commented in Ref. [3] of \cite{nauenberg}). Consistently, it seems fair to nowadays restrict possible criticism to applications indeed using the 1998 version. Unfortunately, all types of applications are criticized in \cite{nauenberg} independently from what particular manner have the author(s) adopted for the energy constraint. An intriguing example of such procedure is the criticism of some 1995-1996 papers (Refs. [19-21] of \cite{nauenberg}) on the possible $q$-generalization of the black-body radiation law. They {\it indeed} satisfy, as they should and as claimed in \cite{nauenberg}, the $T^4$ Stefan-Boltzmann law ({\it explicitly written} in Eq. (18) of Ref. [19] of \cite{nauenberg}). Nevertheless, they do not escape the criticism! It is argued in \cite{nauenberg} that errors have been done in these three papers, and that, if these errors had not been done, the papers would have violated the $T^4$ law, and therefore they also deserve criticism. In addition to this somewhat courageous comment, no remarks are done about the fact that all three were published up to three years {\it before} the need for re-writing the energy constraint became clear.  More significantly, this specific criticism is indeed intriguing since it can be trivially shown that the $T^4$ proportionality law remains the same for {\it all} energy statistical distributions (hence {\it not only} the BG one) as long as the microscopic energy scales linearly with the temperature (i.e., for photons, as long as the distribution depends on the light frequency $\nu$ and the appropriate temperature $T$, {\it only} through $\nu/T$). Only the proportionality coefficient of the $T^4$ law depends on the specific statistics. 
  
The author of \cite{nauenberg} claims to have delivered the epistemological {\it coup de gr\^ace} to nonextensive statistical mechanics. Indeed, expressions like ``unphysical", ``manifestly incorrect", ``devoid of any physical meaning", ``do not have any physical meaning", ``disregarding such basic considerations", ``nonsensical", ``failure of this formalism", ``inconsistencies", ``inconsistent with the fundamental principles of thermodynamics and statistical mechanics",  ``absolutely no physical justification has been given", and analogous ones, have been profusely used in \cite{nauenberg}. We  have essentially argued here that what we are facing is rather the opposite, in the sense that it is precisely the basis of the critique in \cite{nauenberg} which appears to be deeply inconsistent with very many, and by now well established, physical and mathematical facts. We have only addressed the main mispaths and inadvertences in \cite{nauenberg}. There are several more, but the full consideration of them all would demand an appreciable effort which, at the present moment, does not seem worthy. Our overall conclusion is that, although several important and/or interesting points related to nonextensive statistical mechanics still need further clarification, this theory undoubtedly exhibits nowadays a sensible number of physically and mathematically consistent results. Of course, as it has always been, only time will establish its degree of scientific utility in theoretical physics and elsewhere.  \\

\noindent
{\bf ACKNOWLEDGMENTS}\\

M. Nauenberg communicated to me an early version of his arguments prior to publication, and asked for comments. Although he adopted basically none of those which I made on physical grounds, his gentle communication certainly was a polite attitude which I gladly acknowledge here. 
%I also acknowledge L.G. Moyano and F. Baldovin, who kindly allowed me to use the present two figures prior to publication of the full results.    

%\newpage

\begin{figure}
\begin{center}
\includegraphics[width=9cm,angle=0]{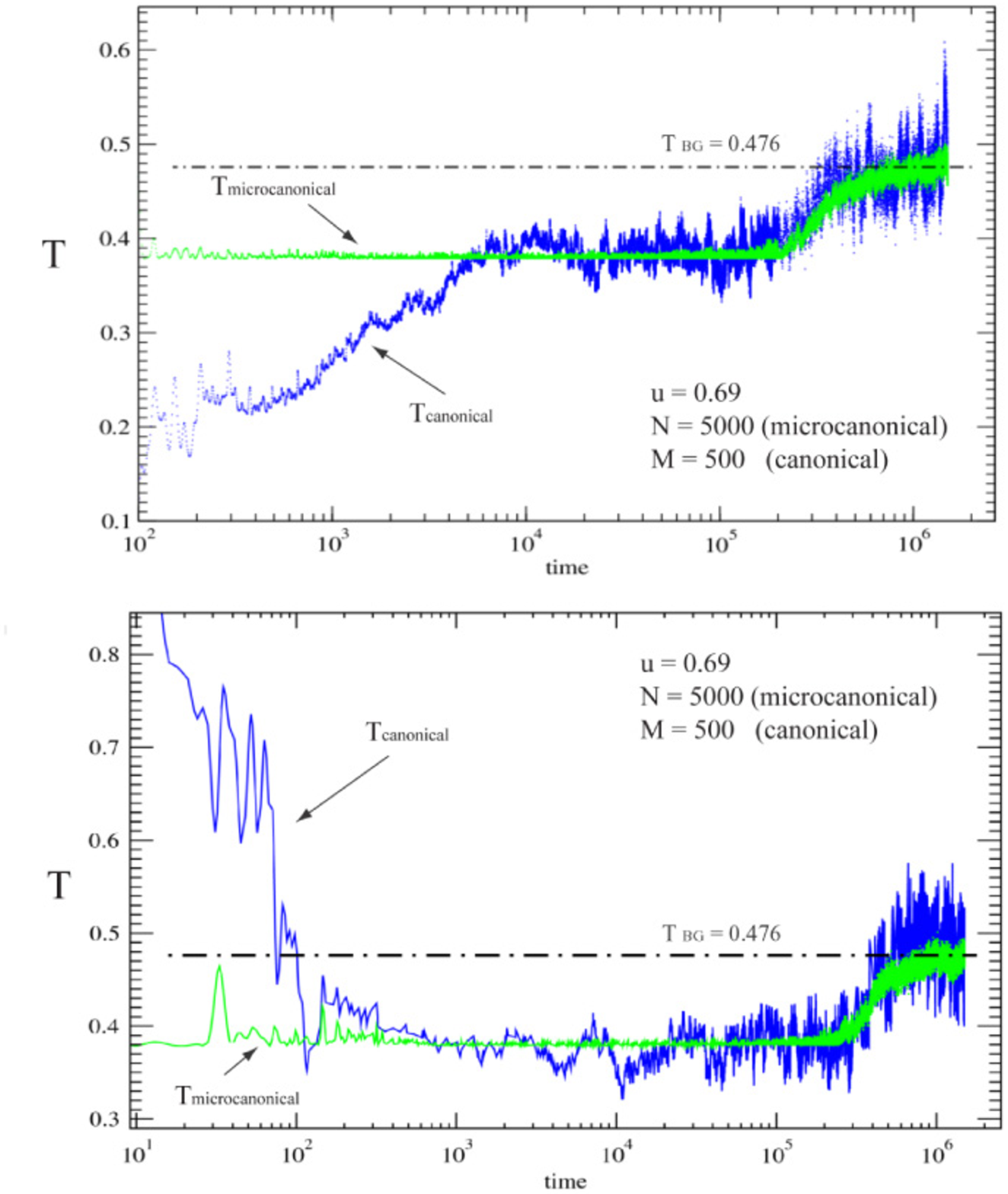}
\end{center}
\caption{\small Time evolution of the temperature $T_{microcanonical}$\\$ \equiv 2 K(N)/N$ ($K(N) \equiv total \;kinetic \;energy$) of one isolated system started with waterbag initial conditions at (conveniently scaled) energy per particle equal to $0.69$ ($N=5000$ rotators; green line), and of the temperature $T_{canonical} \equiv$\\$ 2 K(M)/M$ ($K(M) \equiv subsystem \; total \;kinetic \;energy$)
of a part of it ($M=500$ rotators; blue line). The $M$ rotators were chosen such that their temperature  $T_{canonical}$ 
was initially below (a) or above (b) that of the whole system. It is particularly interesting the fact that, in case (b), the temperature of the subsystem of $M$ rotators crosses the BG temperature $T_{BG}=0.476$ without {\it any} particular detection of it. } 
\end{figure}
\begin{figure}
\begin{center}
\includegraphics[width=8cm,angle=0]{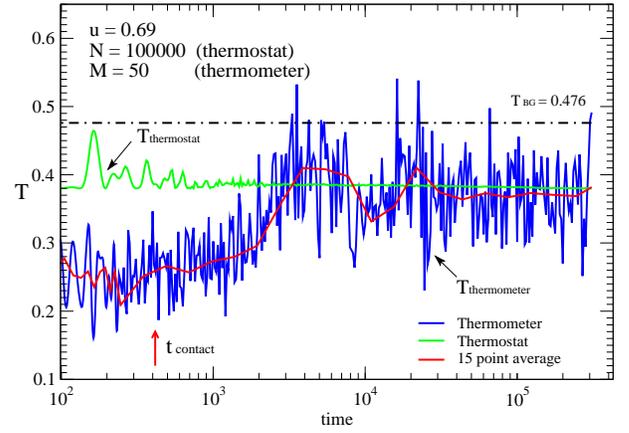}
\end{center}
\caption{\small Time evolution of the temperature $T_{thermostat}\equiv$\\$ 2 K(N)/N$ ($K(N) \equiv thermostat\;total \;kinetic \;energy$) of one infinitely-range-coupled large system (thermostat) started with waterbag initial conditions ($N=100000$ rotators; green line) and of the temperature  $T_{thermometer}\equiv 2 K(M)/M$ ($K(M) \equiv thermometer\;total \;kinetic \;energy$) of one first-neighbor-coupled relatively small system (thermometer) started at Maxwellian equilibrium at a temperature below that of the thermostat ($M=50$ rotators; blue and red lines). The large system is in the quasistationary state (where it is aging!); its (conveniently scaled) energy per particle equals $0.69$. The thermometer-thermostat contact is assured by only one bond per thermometer rotator, and starts at time $t_{contact}$. The intra-thermostat and intra-thermometer coupling constants equal unity; the thermostat-thermometer coupling constant equals $0.001$. The thermalization of the thermometer occurs at the thermostat temperature, and up to time $t=3 \times 10^5$, exhibits no detection of the BG equilibrium temperature $T_{BG}=0.476$ . The same phenomenon with the thermometer initial temperature being {\it larger} than that of the thermostat is not shown, because our numerical results suggest that the $N>>M>>1$ limit has to be satisfied in an even more stringent manner due to the relatively large fluctuations of $T_{thermometer}$. For clarity, not all the points of the curves have been represented, but they have been instead logarithmically decimated.} 
\end{figure}

\end{multicols}


\begin{thebibliography}{99}

\bibitem{nauenberg}M. Nauenberg, Phys. Rev. E {\bf 67}, 036114 (2003). 

\bibitem{buenosaires}First International Conference on Complex Systems in Computational Physics (18-22 October 1993, Buenos Aires). 
Nauenberg  heard, possibly for the first time (in any case, for the first time from me), about the nonextensive formalism in that occasion, where I presented a talk on the subject. Immediately after the Session Chairman opening the talk for discussion, Nauenberg presented to the audience the objections he has now developed in \cite{nauenberg}. 

\bibitem{fisheretal}M.E. Fisher, Arch. Rat. Mech. Anal. {\bf 17}, 377 (1964), J. Chem. Phys. {\bf 42}, 3852 (1965), and J. Math. Phys. {\bf 6}, 1643 (1965); M.E. Fisher and D. Ruelle, J. Math. Phys. {\bf 7}, 260 (1966); M.E. Fisher and J.L. Lebowitz, Commun. Math. Phys. {\bf 19}, 251 (1970). 

\bibitem{rapisarda}V. Latora, A. Rapisarda and C. Tsallis, Phys. Rev. E {\bf 64}, 056134 (2001).

\bibitem{fermi}E. Fermi, {\it Thermodynamics} (1936) [``The entropy of a system composed of several parts is very often equal to the sum of the entropies of all the parts. This is true if the energy of the system is the sum of the energies of all the parts and if the work performed by the system during a transformation is equal to the sum of the amounts of work performed by all the parts. Notice that these conditions are not quite obvious and that in some cases they may not be fulfilled. Thus, for example, in the case of a system composed of two homogeneous substances, it will be possible to express the energy as the sum of the energies of the two substances only if we can neglect the surface energy of the two substances where they are in contact. The surface energy can generally be neglected only if the two substances are not very finely subdivided; otherwise, it can play a considerable role."].   

\bibitem{TMP}C. Tsallis, A.R. Plastino and W.-M. Zheng, Chaos, Solitons and Fractals {\bf 8}, 885 (1997).

\bibitem{uriel}U.M.S. Costa, M.L. Lyra, A.R. Plastino and C. Tsallis, Phys. Rev. E {\bf 56}, 245 (1997).

\bibitem{lyra}M.L. Lyra and C. Tsallis, Phys. Rev. Lett. {\bf 80}, 53 (1998).

\bibitem{lyramoura}F.A.B.F. de Moura, U. Tirnakli and M.L. Lyra, Phys. Rev. E {\bf 62}, 6361 (2000).

\bibitem{ernestogarin}E.P. Borges, C. Tsallis, G.F.J. Ananos and P.M.C. Oliveira, Phys. Rev. Lett. {\bf 89}, 254103 (2002).

\bibitem{fulvioalberto}F. Baldovin and A. Robledo, Europhys. Lett. {\bf 60}, 518 (2002).

\bibitem{fulvioalberto2}F. Baldovin and A. Robledo, Phys. Rev. E {\bf 66}, R045104 (2002).

\bibitem{scienceplastino}A. Plastino, Science {\bf 300}, 250 (2003).

\bibitem{sciencelatora}V. Latora, A. Rapisarda and A. Robledo, Science {\bf 300}, 250 (2003).

\bibitem{LV} A. Provata, G. Nicolis and F. Baras, J. Chem.
Phys. {\bf 110}, 8361 (1999); G. A. Tsekouras and A. Provata,  Phys. Rev. E
{\bf 65}, 016204 (2002).

\bibitem{astero}G.A. Tsekouras, A. Provata and C. Tsallis, cond-mat/0303104.

\bibitem{celia}C. Anteneodo (2003), private communication.

\bibitem{bogho1}B.M. Boghosian, P.J. Love, P.V. Coveney, I.V. Karlin, S. Succi and J. Yepez, cond-mat/0211093 (2002).

\bibitem{bogho2}B.M. Boghosian, P.J. Love and J. Yepez, {\it Galilean-invariant multi-speed entropic lattice Boltzmann models}, preprint (2003).

\bibitem{celiaconstantino}C. Anteneodo and C. Tsallis, Phys. Rev. Lett. {\bf 80}, 5313 (1998).

\bibitem{giansanti}A. Campa, A. Giansanti, D. Moroni and C. Tsallis, Phys. Lett. A {\bf 286}, 251 (2001).

\bibitem{cabral}B.J.C. Cabral and C. Tsallis,, Pys. Rev. E {\bf 66}, 065101(R) (2002).

\bibitem{montemurroetal}M.A. Montemurro, F. Tamarit and C. Anteneodo, Phys. Rev. E {\bf 67}, 031106 (2003); A. Pluchino, V. Latora and A. Rapisarda, cond-mat/0303081.

\bibitem{rapisardarecente}C. Tsallis, A. Rapisarda, V. Latora and F. Baldovin, in {\it Dynamics and Thermodynamics of Systems with Long Range Interactions}, eds. T. Dauxois, S. Ruffo, E. Arimondo, M. Wilkens, Lecture Notes in Physics {\bf 602} (Springer, Berlin, 2002). CONTROLAR 

\bibitem{doye}J.P.K. Doye, Phys. Rev. Lett. {\bf 88}, 238701 (2002).

\bibitem{barabasi}R. Albert and A.L. Barabasi, Phys. Rev. Lett. {\bf 85}, 5234 (2000).

\bibitem{baldovinbrigattitsallis}F. Baldovin, E. Brigatti and C. Tsallis, cond-mat/0302559. 

\bibitem{giansantietal}A. Campa, A. Giansanti and D. Moroni,  Physica A {\bf 305}, 137 (2002).

\bibitem{sokal}A.C.D. van Enter, R. Fernandez and A.D. Sokal, J. Stat. Phys. {\bf 72}, 879 (1993). 

\bibitem{einstein}A. Einstein, Annalen der Physik {\bf 33}, 1275 (1910) [
``Usually $W$ is put equal to the number of complexions... In order to
calculate $W$, one needs a {\it complete} (molecular-mechanical) theory of the
system under consideration. Therefore it is dubious whether the Boltzmann
principle has any meaning without a complete molecular-mechanical theory or
some other theory which describes the elementary processes.
$S=\frac{R}{\cal N}\log W+\;{\rm const}.$ seems
without content, from a phenomenological point of view, without giving in
addition such an {\it Elementartheorie}.'' (Translation: Abraham Pais, {\it
Subtle is the Lord...}, Oxford University Press, 1982)]. 

\bibitem{proof1}S. Abe and A.K. Rajagopal, J. Phys. A {\bf 33}, 8733 (2000).

\bibitem{proof2}S. Abe and A.K. Rajagopal, Europhys. Lett. {\bf 52}, 610 (2000).

\bibitem{proof3}S. Abe and A.K. Rajagopal, Phys. Lett. A {\bf 272}, 341 (2000), and Europhys. Lett. {\bf 55}, 6 (2001).

\bibitem{scienceabe}S. Abe and A.K. Rajagopal, Science {\bf 300}, 249 (2003).

\bibitem{theorem1}R.J.V. Santos, J. Math. Phys. {\bf 38}, 4104 (1997).

\bibitem{theorem2}S. Abe, Phys. Lett. A {\bf 271}, 74 (2000).

\bibitem{stability}S. Abe, Phys. Rev. E {\bf 66}, 046134 (2002).

\bibitem{production}V. Latora, M. Baranger, A. Rapisarda and C. Tsallis, Phys. Lett. A {\bf 273}, 97 (2000). See also the $q$-generalization of Pesin theorem, conjectured in \cite{TMP}, and recently proved by F. Baldovin and A. Robledo, cond-mat/0304410. 

\bibitem{moyanobaldovin}L.G. Moyano, F. Baldovin and C. Tsallis, (preprint) (2003), to be published. 

\bibitem{multiplicative}C. Anteneodo and C. Tsallis, cond-mat/0205314.

\bibitem{nonlinear}A.R. Plastino and A. Plastino, Physica A  {\bf 222}, 347 (1995); C. Tsallis and D.J. Bukman, Phys. Rev. E {\bf 54}, R2197 (1996).

\bibitem{escort}C. Beck and F. Schlogl, {\it Thermodynamics of Chaotic Systems} (Cambridge University Press, Cambridge, 1993).

\bibitem{tsallissouza}C. Tsallis and A.M.C. Souza, Phys. Rev. E {\bf 67}, 026106 (2003). 

\bibitem{beckcohen}C. Beck and E.G.D. Cohen, Physica A {\bf 321} (2003).

\bibitem{souzatsallis}A.M.C. Souza and C. Tsallis, cond-mat/0301304.

\end{thebibliography}
\end{document}